\documentstyle[aps]{revtex}
\begin{document}
\draft
\title{Lattice vibrations of $\alpha'$-NaV$_2$O$_5$ }
\author{M. N. Popova, A. B. Sushkov, S. A. Golubchik, B. N. Mavrin,
        and V. N. Denisov}
\address{Institute of Spectroscopy, Russian Academy of Sciences,\\
         142092 Troitsk, Moscow reg., Russia}
\author{B. Z. Malkin and A. I. Iskhakova}
\address{Kazan State University, 420008 Kazan, Russia}
\author{M. Isobe and Y. Ueda}
\address{Institute for Solid State Physics, The University of Tokyo \\
         7--22--1 Roppongi, Minato-ku, Tokyo 106, Japan}
\date{\today}
\maketitle

\begin{abstract}
We have measured far infrared reflectance and transmittance spectra
as well as Raman scattering spectra of $\alpha'$-NaV$_2$O$_5$ single
crystals for all the principal polarizations. The temperature range
above the phase transition temperature $T_c=35$~K was investigated,
mainly. On the basis of this experimental study and of the lattice
dynamics calculations we conclude that the symmetry of NaV$_2$O$_5$
in the high temperature phase is described by the centrosymmetric
$D_{2h}^{13}$ space group.  The assignment of the observed phonons is
given.  Values of dielectric constants are obtained from the infrared
data.  Asymmetric shapes of several infrared lines  as well as higher
order infrared vibrational spectra are discussed. The crystal field
energy levels of the 3$d$ electron localized at the V$^{4+}$ site
have been calculated in the framework of the exchange charge model
using the values of effective charges obtained from the lattice
dynamics calculations.  According to the results of these
calculations, the earlier observed broad optical bands in the region
of 1~eV can be interpreted as phonon assisted $d-d$ transitions.
\end{abstract}

\section{Introduction}

The vanadate $\alpha'$-NaV$_2$O$_5$ has
attracted a considerable interest recently as the second
inorganic compound undergoing the spin-Peierls
transition (at the highest known temperature for the
spin-Peierls compounds, $T_c=35$~K~\cite{c1}). The
spin-Peierls transition is expected to occur within a
system of linear spin-1/2 Heisenberg antiferromagnetic
chains coupled to a three-dimensional phonon field. As a
result of such a coupling, magnetic atoms of the chain
dimerize and a spin gap opens~\cite{c2}. One-dimensional
magnetic properties of NaV$_2$O$_5$ above 35~K follow
from magnetic susceptibility~\cite{c1},
ESR~\cite{hemberger} and
angle-resolved photoemission~\cite{kobayashi} measurements.

Below 35 K the lattice dimerizes, as observed by
X-ray~\cite{c3} and Raman~\cite{c4,c5} scattering,
infrared transmission~\cite{LJETP} and reflection~\cite{d.smirnov}
measurements, while the magnetic
susceptibility decreases isotropically, thus showing a
spin gap formation~\cite{c4}. The size of the gap
$\Delta=10$~meV follows from inelastic neutron
scattering study of NaV$_2$O$_5$ single
crystals~\cite{c3,yoshihama2}.

The structure of NaV$_2$O$_5$ contains double chains of
edge sharing distorted VO$_5$ pyramids running along the
orthorhombic $b$-axis (see Fig.~1). These double chains
are linked together via common corners of the pyramids
to form the $ab$-layers.
Na atoms lie between the layers~\cite{carpy,c1}.
The structure of NaV$_2$O$_5$ looks like the structure of
V$_2$O$_5$~\cite{backman} intercalated with sodium.
In an early X-ray room
temperature investigation on polycrystalline samples of
NaV$_2$O$_5$ Carpy and Galy~\cite{carpy} suggested the
noncentrosymmetric space group $C_{2v}^7$-$P2_1mn$
with two nonequivalent vanadium positions in the unit
cell. The picture of magnetic chains of V$^{4+}$O$_5$
($S=1/2$) pyramids isolated by nonmagnetic chains of
V$^{5+}$O$_5$ ($S=0$) pyramids proposed to account for
one-dimensional magnetic properties of this mixed
valence (V$^{4.5+}$) compound is compatible with this
space group~\cite{c1}.

However, the recent redetermination of the structure by single
crystal X-ray diffraction at room temperature was in favor of the
centrosymmetric $D_{2h}^{13}$-$Pmmn$ group with only one vanadium
position in the structure~\cite{smolinski,meetsma}.  Though the
topology of the structure remains essentially the same as in the
previously proposed noncentrosymmetric space group~\cite{carpy}, the
possibility for charge ordering is, however, lost in the new higher
symmetry group.
Smolinski at al.~\cite{smolinski} and Horsch and
Mack~\cite{horsch} suggested a quarter-filled ladder
model for NaV$_2$O$_5$, with the spins carried by
V--O--V molecular orbitals on the rungs of the ladder.
They argued that the exchange interaction along the
ladder is much greater than that between the neighboring
ladders which would explain the one-dimensional magnetic
properties of the high temperature phase of NaV$_2$O$_5$.
The transition at 35~K was supposed to be an
ordinary spin-Peierls transition. Quite recent
$^{51}$V-NMR experiment on a single-crystalline sample
of NaV$_2$O$_5$ also revealed only one vanadium position
in the high temperature phase but pointed unambiguously
to the existence of two different vanadium sites
occupied by V$^{4+}$ and V$^{5+}$ at liquid helium
temperatures~\cite{ohama}. Thus, the transition at 35~K
is connected with a structure and a charge ordering
processes. Very recently, Seo and Fukuyama~\cite{seo}
and Mostovoy and Khomskii~\cite{mostovoy} proposed a
zig-zag scheme of V$^{4+}$--V$^{5+}$ ordering. Seo and
Fukuyama argued that, as a result, two-dimensional
lattice of antiferromagnetic dimers is formed~\cite{seo},
while Mostovoy and Khomskii gave reasons in
support of a system of alternating
chains~\cite{mostovoy}. Thalmeier and Fulde \cite{Fulde}
have presented some theoretical reasons for the primary
charge ordering which provides neighboring linear
V$^{4+}$ and V$^{5+}$ chains with the subsequent spin-Peierls
transition. Two close transitions near 35 K in
NaV$_2$O$_5$ were detected by K\"{o}ppen et al. by the
thermal expansion measurements \cite{Koppen}.

In view of these recent works, the symmetry problem of the high
temperature phase seems to be a matter of great urgency. Raman and
infrared measurements could give an additional information to clarify
the question whether the space group is centrosymmetric or not,
because of totally different selection rules in these two cases. We
reanalyzed our earlier infrared and Raman spectra of NaV$_2$O$_5$
\cite{we} and found that they are in a better agreement with the 
centrosymmetric $D_{2h}^{13}$ space group than with the
noncentrosymmetric $C_{2v}^7$ one.  However, in our work~\cite{we} we
did not measure infrared spectra in the ${\bf E}||{\bf c}$
polarization.  Also, the signal-to-noise ratio of Raman spectra was
rather low. In the present work, we reinvestigate vibrational spectra
of the high temperature phase of NaV$_2$O$_5$ using different single
crystals, including extraordinary big ones. We present far-infrared
reflectivity as well as Raman-scattering spectra for all principal
polarizations. In addition, transmittance spectra were studied. We
show that our results are in a much better agreement with the
centrosymmetric $D_{2h}^{13}$ group than with the noncentrosymmetric
$C_{2v}^7$ one.  The assignment of vibrational modes is given based
on a comparison with results on the previously studied
V$_2$O$_5$~\cite{abello} and on lattice dynamics calculations of this
work performed in the framework of the rigid ion model.

\section{Experimental}

Single crystals of stoichiometric
$\alpha'$-NaV$_2$O$_5$ used in this study were
grown by a melt growth method using NaVO$_3$ as a flux
\cite{isobe kagami ueda}. Samples from different
batches were used. One sample was $1.3\times 8\times
1$~mm, another one was $3\times 17.3\times 1.6$~mm along
$a$-, $b$- and $c$-axes, respectively. For transmission
measurements we have prepared four thin samples cleaved
perpendicular to the $c$-axis. Their thicknesses were
$110\pm1$, $45\pm5$, $14\pm1.5$ and $6\pm1$~$\mu$m. The
samples were checked with X-ray diffraction,
magnetization and ESR measurements. They exhibited a
sharp transition at about 35~K.

Reflection and transmission measurements were performed
with a BOMEM DA3.002 Fourier-transform spectrometer at
nearly normal incidence of polarized infrared radiation.
The following geometries of the experiment were used: 1)
${\bf k}||{\bf c}$, ${\bf E}||{\bf a}$ and ${\bf
E}||{\bf b}$; 2) ${\bf k}|| {\bf a}$, ${\bf E}||{\bf
c}$ and ${\bf E}||{\bf b}$. Room temperature reflectance
and transmittance spectra were measured in a spectral
range 30--5000~cm$^{-1}$ with a resolution
0.5--2.0~cm$^{-1}$. Using both reflectance and
transmittance spectra, the absorption coefficient
$\alpha$ was calculated. Low temperature (down to 6~K)
transmittance spectra were measured with a He vapor
cryostat in the spectral range 30--1000~cm$^{-1}$ with a
resolution 0.05--1.0~cm$^{-1}$.

Raman spectra were excited at room temperature by the
514~nm and 488~nm lines of an Ar-ion laser in
backscattering geometries, dispersed by a home-made
triple spectrograph and recorded using a multichannel
system consisting of an image intensifier tube with a
multichannel plate and a vidicon.

\section{Results}


\subsection{Factor-group analysis}

There are two formula units and, hence, 16 atoms in the
NaV$_2$O$_5$ orthorhombic unit cell with lattice
constants $a=1.1316$~nm, $b=0.3611$~nm,
$c=0.4797$~nm~\cite{carpy,smolinski,meetsma}. Below, we
present the results of factor-group analysis for both
centrosymmetric $D_{2h}^{13}$~\cite{smolinski,meetsma}
and noncentrosymmetric $C_{2v}^7$~\cite{carpy} space
groups.

\paragraph{Space group $D_{2h}^{13}$-$Pmmn$}

The notation $Pmmn$ refers to the standard setting, such
that $x||a$, $y||b$, $z||c$. It follows from X-ray
diffraction data of Refs.~\cite{smolinski,meetsma} that
Na atoms occupy $2b$ positions (the corresponding
fractional atomic coordinates are defined by the basis
vectors ${\bf r}_1($ Na $)=-{\bf
r}_2($Na$)=(1/4,-1/4,z_1),$ $z_1=0.8592$) and oxygen O1
atoms occupy $2a$ positions (${\bf r}_1($O1$)=-{\bf
r}_2($O1$)=(1/4,1/4,z_2),$ $z_2=0.5195$), both these
positions having $C_{2v}^z$ local symmetry. V, O2 and O3
atoms reside in different $4f$ positions (${\bf
r}_1($A$)=-{\bf r}_3($A$)=(x_A,1/4,z_A),$ ${\bf
r}_2($A$)=-{\bf r}_4($A$)=(1/2-x_A,1/4,z_A),$
$x_V=0.40212,$ $z_V=0.39219,$ $x_{O2}=0.57302,$
$z_{O2}=0.48769,$ $x_{O3}=0.38548,$ $z_{O3}=0.05803$)
with the local symmetry $C_s^{xz}$. These positions
yield the following irreducible 
representations~\cite{mavrin,rousseau}: 
\[ C_{2v}^z:
\Gamma=A_g+B_{2g}+B_{3g}+B_{1u}+B_{2u}+B_{3u} \] 
\[ C_s^{xz}:
\Gamma=2A_g+B_{1g}+2B_{2g}+B_{3g}+A_u+2B_{1u}+B_{2u}+2B_{3u} \] 
Multiplying the representations given above by
the number of different positions of the appropriate
symmetry, summarizing them and subtracting acoustic
modes ($B_{1u}+B_{2u}+B_{3u}$), we obtain the following
NaV$_2$O$_5$ optical vibrational modes: 
\begin{eqnarray}
\Gamma_{NaV_2O_5}^{vib}(Pmmn)=8A_g(aa,bb,cc)+3B_{1g}(ab)+
 \nonumber \\
+8B_{2g}(ac)+5B_{3g}(bc)+3A_u+
 \nonumber \\
+7B_{1u}({\bf E}||{\bf c})+4B_{2u}({\bf E}||{\bf b})+7B_{3u}({\bf E}||{\bf a}) 
\end{eqnarray}
There are 45 vibrational modes in total. $A_u$ modes
being silent, 24 Raman ($A_g$, $B_{1g}$, $B_{2g}$,
$B_{3g}$) and 18 infrared ($B_{1u}$, $B_{2u}$, $B_{3u}$)
active modes are expected to be found in the
spectra of NaV$_2$O$_5$, provided the crystal space
group is $D_{2h}^{13}$.

\paragraph{Space group $C_{2v}^7$-$P2_1mn$}

In their original work~\cite{carpy} Carpy and Galy
adopted the axes setting for the $P2_1mn$ space group.
Below, we use the standard setting for the $Pmn2_1$
space group: $x||b$, $y||c$, $z||a$. There are two
nonequivalent V positions, five nonequivalent O
positions and one Na position in this group, all of them
being $2a$ positions with $C_s^{yz}$ local symmetry. In
the same way as in the previous case, using
tables~\cite{rousseau} and subtracting acoustic modes
($A_1+B_2+B_1$), we find the following vibrational
modes: 
\begin{eqnarray}
\Gamma_{NaV_2O_5}^{vib}(Pmn2_1)=15A_1(aa,bb,cc;{\bf E}||{\bf a})+
\nonumber \\
+8A_2(bc)+7B_1(ab;{\bf E}||{\bf b})+ 15B_2(ac;{\bf E}||{\bf c}) 
\end{eqnarray}
There are 45 optical modes again. But in the case of
this noncentrosymmetric space group all of them are
Raman active, 37 of them are also infrared active.


\subsection{Infrared spectra}

Fig. 2 shows the room temperature far-infrared
reflectivity spectra of NaV$_2 $O$_5$ for different
polarizations of the incident light. Experimental data
are presented by open circles. Measured spectra were
least-squares fitted by the spectra computed according
to the expression \begin{equation} {\cal
R}=\left|\frac{\sqrt{\varepsilon}-1}
{\sqrt{\varepsilon}+1}\right|^2 \end{equation} The
classical dispersion formula for N independent damped
oscillators was used: \begin{equation}
\varepsilon=\varepsilon_{\infty}+\sum_{i=1}^N \frac{4\pi
f_i \omega_i^2} {\omega_i^2-\omega^2-\imath \gamma_i
\omega}  \label{eps} \end{equation} For ${\bf E}||{\bf
b}$ and ${\bf E}||{\bf a}$ polarizations the number of
oscillators and initial values of parameters were taken
from the transmittance spectra (\cite{we} and the
present work). The peculiarity at 1014~cm$^{-1}$ in
${\bf E}||{\bf a}$ polarization crossed out in Figs.~2
and~3 and observed also in ${\bf E}||{\bf b}$
polarization for some samples depends on a particular
sample and is not, evidently, an intrinsic property of
NaV$_2$O$_5$. It was not taken into account in the
fitting procedure. In addition to weakly damped phonon
oscillators, an overdamped oscillator centered at about
300~cm$^{-1}$ ($\omega_i=291$~cm$^{-1}$,
$\gamma_i=260$~cm$^{-1}$, $f_i=0.38$) was introduced
in ${\bf E}||{\bf a}$ polarization to account for a low
frequency part of a broad absorption band of a complex
two-humped shape found in our previous study~\cite{we}
(see also Fig~3). We failed to model the high frequency
hump of this band centered at about 1000~cm$^{-1}$ by a
similar oscillator and did not try to use a more
complicated model. This results in a not so good fitting
of the high frequency part of the reflectance spectrum.
In ${\bf E}||{\bf a}$ polarization the phonon at about
150~cm$^{-1}$ could not be fitted well. This line is
strongly asymmetric in transmittance spectra, obviously,
due to the interaction with the underlying broad band.

A small bump in reflection at 939~cm$^{-1}$ shown by the
arrow in Fig~2 corresponds to the Fano-type resonance
\cite{fano} well seen in the absorbance spectrum
(Fig.~3). One more such a resonance becomes visible
below 200~K at the frequency of about 91~cm$^{-1}$
(see Fig.~4 and also Ref.~\cite{LJETP}). 
We fitted the absorption coefficient in
the vicinity of these two strongly asymmetric lines by
the expression~\cite{fano}: 
\begin{equation}
\alpha(\omega)=\alpha_B(\omega)+\alpha_0\frac{q^2+2\xi q
-1}{1+\xi^2} 
\label{alpha} 
\end{equation} 
where
$\xi=(\omega-\omega_r)/\gamma$, \, $\alpha_B(\omega)$ is
a slowly varying broad band absorption (it is shown by
a dashed line in the vicinity of the 939~cm$^{-1}$
sharp resonance in Fig.~3), $\alpha_0$, $\omega_r$,
$\gamma$ and $q$ being variable parameters. Such an
expression describes various physical situations of a
sharp transitions being overlapped by a broad continuum.
The lineshape of a sharp transition is changed by an
interference with a continuum and depends essentially on
the strength of an interaction between discrete and
continuum states. The parameter $q$ being inversely
proportional to the matrix element of an interaction,
the case $|q|=\infty$ corresponds to zero interaction
and results in a normal Lorenzian resonance, $|q|=1$
yields a dispersion-like curve, while $|q|=0$ gives an
inverted Lorenzian (antiresonance).
The ratio $\alpha_0/\alpha_B$ shows what fraction of the continuum
states interacts with a sharp excited state. The results
of the fitting are displayed in the inset of Fig.~3 and
in Fig.~4. A similar fitting should be performed for the
resonance at about 150~cm$^{-1}$ but we failed to
construct $\alpha_B(\omega) $ in this case. The
parameters obtained from the fittings are listed in
Table~1. $\omega_{TO}$ and $\gamma_{TO}$ stand for
$\omega_i$ and, correspondingly, $\gamma_i$ of 
Eq.~(\ref{eps}) or for $\omega_r$ and $\gamma$ of
Eq.~(\ref{alpha}). LO frequencies and damping constants
were calculated as complex roots of the equation
$\varepsilon(\omega)=0$.

The left inset of Fig. 4 presents the temperature dependence of the
Fano parameter $q$ for the spectral line near 91~cm$^{-1}$ for the
temperatures higher than $T_c=35$~K. It should be mentioned that
below 35~K, the shape of this line changes to an ordinary Lorenzian
lineshape (see the right inset of Fig.~4 and also \cite{LJETP}).
Simultaneously, a continuum absorption diminishes markedly in this
spectral region while it practically does not change at the maximum
of the low-frequency hump at 320~cm$^{-1}$.

With decreasing the temperature, besides the asymmetric
resonance at 91~cm$^{-1}$ in ${\bf E}||{\bf a}$
transmission, two lines at 215 and 225~cm$^{-1}$ appear
in ${\bf E}||{\bf b}$ transmittance spectra as well
\cite{LJETP,we}. We have studied the resonances at 91
and 939~cm$^{-1}$ (${\bf E}||{\bf a}$); 215 and
225~cm$^{-1}$ (${\bf E}||{\bf b}$) for
the samples of different thicknesses and found that
while the intensity of 91, 939 and 225~cm$^{-1}$ lines
grow proportionally to the sample thickness $d$ (that
is, $\alpha=const$), the intensity of the 215~cm$^{-1}$
line does not, practically, depend on the thickness
($\alpha d \simeq const$). Consequently, while the
frequencies 91 and 939~cm$^{-1}$ (${\bf E}||{\bf a}$)
and 225~cm$^{-1}$ (${\bf E}||{\bf b}$) correspond to
intrinsic resonances, $\omega=215$~cm$^{-1}$ must refer
to a surface excitation. All the observed infrared
phonon frequencies together with the calculated ones are
displayed in Table~1.

NaV$_2$O$_5$ crystals are well transparent in the
frequency region between 2500 and 4500~cm$^{-1}$ and
below 100~cm$^{-1}$. In these regions, an interference
pattern was observed in (${\bf E}||{\bf a}$) and (${\bf
E}||{\bf b}$) transmittance spectra of the samples of
good quality. We also managed to observe the
interference pattern below 100~cm$^{-1}$ in ${\bf E}||{\bf c}$
transmittance of 1.3~mm thick sample. By
measuring the distances $\Delta$ between the
interference maxima, we have found the values of the
refractive indexes $n$ according to the relation
\begin{equation} 
\Delta=\frac1{2dn},  
\label{Delta}
\end{equation} 
$d$ being the thickness of a sample. The appropriate values of
$\varepsilon=n^2$ are listed in Table~1.

We also looked for the higher order vibrational spectra
by measuring the transmittance of thick
($d$=0.4--3.0~mm) samples in the frequency range
1000--4000~cm$^{-1}$. While no pronounced features were
found in ${\bf E}||{\bf a}$ and ${\bf E}||{\bf b}$
polarizations, sharp resonances were observed in ${\bf
E}||{\bf c}$ (${\bf k}||{\bf a}$) polarization at 1930,
2858 and, possibly, 1072 and 1270~cm$^{-1}$, the latter
two lines being somewhat masked by the edge of a strong
phonon at 955~cm$^{-1}$ (see Fig.~5).

\subsection{Raman spectra}

Polarized room-temperature Raman spectra of NaV$_2$O$_5$
in the spectral range 80--1000~cm$^{-1}$ are shown in
Fig.~6. One can see immediately that the three diagonal
components $aa$, $bb$, $cc$ of the Raman scattering
tensor differ markedly one from another which points to
a considerable anisotropy of the structure. The most
intense spectra were observed in the $A_g$ geometry
$a(cc)\bar{a}$. The intensity of the lines marked by
asterisks in $B_{ig} (i=1, 2, 3)$ spectra depended 
strongly on slight variations in the
sample orientation. Evidently, these lines are present
due to a leakage of strong lines from $A_g$ geometries.
We failed to assign for certain a weak feature near
100~cm$^{-1}$ in the $b(ac)\bar{b}$ spectrum overlapped
by a strong unshifted line present in this geometry.
Possibly, it comes from a leakage of a very strong line
90~cm$^{-1}$ from the $(cc)$ polarization. Frequencies
of the observed Raman modes together with the calculated
ones are collected in Table~2.

As we have already communicated~\cite{we}, besides
relatively narrow lines, a broad band with a maximum
near 600~cm$^{-1}$ is observed in the $c(aa)\bar{c}$
spectrum (see Fig.~6). Since this band appears under
both 514.5~nm and 488~nm excitation, we conclude that it
originates from the Raman scattering process. However, a
big width of this band (213~cm$^{-1}$), practically
independent of the temperature, does not permit to
assign it to fundamental modes.

We have also studied Raman spectra of Na-deficient
samples Na$_{1-x}$V$_2$O$_5$ ($x=0, 0.05, 0.10,
0.15$). The most prominent changes occur in the $A_g$
($aa$) spectrum (see Fig.~7). The 447~cm$^{-1}$ Raman
line moves to higher frequencies with growing $x$. Its
position shown by the vertical dashed lines in Fig.~7 is
477~cm$^{-1}$ for the sample with $x=0.15$. The maximum
of the broad band moves in the opposite direction,
namely, from 632~cm$^{-1}$ for $x=0$ to 562~cm$^{-1}$
for $x=0.15$. Such a change of the frequency difference
between these two Raman bands is caused, probably, by a
change of intermode interaction. The shape of the broad
band can be approximated well by a Gaussian for all the
values of $x$, its width growing from 213~cm$^{-1}$ for
$x=0$ to 290~cm$^{-1}$ for $x=0.15$. As for phonon Raman
lines, their shape is almost Lorenzian, their width
grows too. For example, the lines 177, 301 and
531~cm$^{-1}$ broaden from, correspondingly, 11, 18 and
20~cm$^{-1}$ for $x=0$ to 16, 27 and 34~cm$^{-1}$ for
$x=0.15$. The broadening of Raman bands with growing $x$
is, evidently, connected with an increase of lattice
disorder.

It is difficult to compare absolute intensities of the
spectra for different $x$. However, certain conclusions
concerning the relative intensities in a given spectrum
can be drawn. The most prominent features are the growth
of the intensity of the 301~cm$^{-1}$ line and the
appearance of a new line at 988~cm$^{-1}$ for $x=0.15$.
All these results were obtained by decomposing of the
observed spectrum into separate spectral contours. An
example of such a decomposition is shown in Fig.~8.

\subsection{Calculations of vibrational spectra}

To obtain an information about the phonon spectrum of
NaV$_2$O$_5$ throughout the Brillouin zone which is
necessary for the analysis of the spin-phonon
interaction effects, we have considered the lattice
dynamics of this crystal in the framework of the rigid
ion model. The goal of this study is to display the
basic pairwise interion interactions which determine the
main features of the measured Raman and infrared
transmittance and reflectance spectra.

From the measured large TO--LO splittings of some normal
modes at the Brillouin zone center ($\Gamma$ point), it
is clear that long-range Coulomb forces play an
essential role in the formation of the vibrational
spectrum of NaV$_2$O$_5$. The potential energy of the
lattice was represented by a sum of Coulomb and
non-Coulomb interactions. The Coulomb terms in the
dynamical matrix were calculated exactly by using the
Ewald method. Non-Coulomb interactions in the form of
the Born-Mayer potentials with the exponential
dependence on the interionic distance $r$ ($\varphi
_{ij}(r)=A_{ij}\exp (-r/\rho_{ij})$) were introduced
between V-O (five bonds per vanadium ion), Na--O (eight
bonds per sodium ion) and O--O neighboring ions at
interionic distances shorter than 0.325~nm. Because of
the nonequivalence of the oxygen O1, O2 and O3 ions we
should introduce different potentials for different
types of bonding. At the initial step we confined
ourselves by only 10 fitting parameters (instead of 34
independent force constants for the V$_2$O$_5$ lattice
in Ref.\cite{abello}) including ion charges $Z(A)$ (a
condition of the lattice neutrality brings the relation
$Z($Na$)+Z($O1$)+2Z($V$)+2Z($O2$)+2Z($O3$)=0$ about) and
$A_{ij},$ $\rho_{ij}$ constants for V--O, Na--O and
O--O pairs of ions.

The theoretical analysis of the vibrational spectra has
been carried out for both proposed in the literature
lattice structures. We did not obtain any physically
well-grounded set of parameters which might provide the
stable $C_{2v}^7$ lattice structure. So, we discuss
further in this section only vibrations of the
centrosymmetric lattice with the $D_{2h}^{13}$ space
group.

The orthogonal transformation of the atomic
displacements to the symmetrized and normalized linear
combinations, namely,

$u_\alpha (\Gamma _{1u},A)=(u_{1\alpha }(A)+u_{2\alpha
}(A)+u_{3\alpha} (A)+u_{4\alpha }(A))/2,$

$u_\alpha (\Gamma _{1g},A)=(-u_{1\alpha }(A)-u_{2\alpha
}(A)+u_{3\alpha }(A)+u_{4\alpha }(A))/2,$

$u_\alpha (\Gamma _{2u},A)=(-u_{1\alpha }(A)+u_{2\alpha
}(A)-u_{3\alpha}(A)+ u_{4\alpha }(A))/2,$

$u_\alpha (\Gamma _{2g},A)=(u_{1\alpha }(A)-u_{2\alpha
}(A)-u_{3\alpha}(A)+ u_{4\alpha }(A))/2,$

$u_\alpha (\Gamma _{1u},B)=(u_{1\alpha }(B)+u_{2\alpha
}(B))/\sqrt{2},$

$u_\alpha (\Gamma _{1g},B)=(-u_{1\alpha }(B)+u_{2\alpha
}(B))/\sqrt{2},$\\where $A$ stands for V, O2 or O3
ions, and $B$ --- for Na or O1 ions, divides the
dynamical matrix at the $\Gamma$-point into blocks
corresponding to the irreducible representations of the
crystal factor-group. Here,

for $\alpha =x,$ $\Gamma _{1u}=B_{3u},$ $\Gamma
_{1g}=B_{2g},$ $\Gamma_{2u}=B_{1u},$ $\Gamma
_{2g}=A_g,$

for $\alpha =y,$ $\Gamma_{1u}=B_{2u},$ $\Gamma
_{1g}=B_{3g},$ $\Gamma_{2u}=A_u,$
$\Gamma_{2g}=B_{1g},$

for $\alpha =z,$ $\Gamma _{1u}=B_{1u},$ $\Gamma
_{1g}=A_g,$ $\Gamma_{2u}=B_{3u},$
$\Gamma_{2g}=B_{2g}.$

From fitting of the calculated eigenvalues of the
dynamical matrix to the measured frequencies of the
lattice normal modes, we obtained the following values
of effective ionic charges: $Z($V$)=2.405,$
$Z($Na$)=0.83,$ $Z($O1$)=-1.22,$ $Z($O2$)=-1.23,$
$Z($O3$)=-0.98$ in units of the proton charge. The
apical oxygen ion O3 closest to the vanadium ion has the
lowest charge. The average values of parameters defining
the Born-Mayer potentials equal $A_{ij}=12.2;$ 0.82;
12.8 (in units of 10$^3$ eV), $\rho _{ij}=$0.0181; 0.031
and 0.0212~nm for the V--O, Na--O and O--O bonds,
respectively. It should be noted that we had to correct
some force constants calculated with these potentials
(the relative values of corrections are the most
significant for the distant O--O bonds but do not exceed
50\%) to improve a comparison between theory and
experiment.

The calculated frequencies of the lattice normal modes
at the Brillouin zone center are presented in Tables 1
and 2 for infrared and, correspondingly, Raman active
modes. The calculated frequencies of silent $A_u$ modes
equal 120, 167 and 572~cm$^{-1}$. Acoustical properties
of the lattice are defined by nine elastic constants,
the predicted values of $C_{11}=17.7,$ $C_{12}=9.7,$
$C_{22}=23.6$ (in units of 10$^{10}$ N/m$^2$) are less
dependent on variations of the model parameters.

The measured components of the high frequency dielectric tensor
$\varepsilon_\infty$ remarkably differ from unity (see Table~1), thus
a neglect of the electronic polarization is a very crude
approximation in this case, in particular, when estimating LO--TO
splittings in the $\Gamma$ point. However, for most of the infrared
active normal modes, our model gives a satisfactory description of
the longitudinal macroscopic electric field induced by the vibrations
of ions.  A very strong damping of the $B_{1u}$ TO mode at the
frequency of 591~cm$^{-1}$ may be the reason for a large difference
between the calculated and measured LO--TO splitting (see Table 1);
in the case of the $B_{2u}$ TO mode at 365~cm$^{-1}$ our model gives
strongly overestimated frequency of the corresponding LO mode.
Rather large discrepancies between several calculated frequencies of
the Raman active modes and the experimental data (see Table~2)
clearly demonstrate that some significant interactions, in
particular, the three body forces, greatly affecting frequencies of
the bending vibrations, are to be included in the more thorough study
of the lattice dynamics of this system.

May be, the most interesting result of this analysis of
the NaV$_2$O$_5$ lattice dynamics is the predicted soft
mode behavior of the transverse acoustical mode at the
Brillouin zone boundary (with the wave vector
${\bf k}_0=\pi (0,0,1/c)$) polarized in the $ac$-plane. Due to a
competition of the long range Coulomb and short range
non-Coulomb forces, the corresponding branch of the
vibrational spectrum moves to the range of imaginary
frequencies when approaching the ${\bf k}_0$ point thus
opening a possibility to consider the NaV$_2$O$_5$
crystal as an improper virtual ferroelastic. To
stabilize the lattice in respect to ${\bf k}_0$-excitations,
we had to introduce an attractive
interaction between the neighboring V$_1$ and V$_2$
(V$_3$ and V$_4$) ions along the $a$-axis with the
significant bending force constant of approximately 5
N/m. The charge ordering in the subsystem of V ions can
destroy a balance between the forces of different signs
and induce a freezing of the soft mode atomic
displacements (the unit cell doubles in the $c$
direction, the neighboring layers shift in the opposite
directions, in each layer the right and left legs of the
vanadium ladders become nonequivalent due to shifts of
V$_1$--V$_2$ and V$_3$--V$_4$ rungs along opposite
directions in the $ac$-plane) as a precursor of the
subsequent magnetic ordering with doubling of a unit
cell in $a$-, $b$- and $c$-directions.

\section{Discussion}

\subsection{Symmetry group of NaV$_2$O$_5$}

Table~3 summarizes the observed vibrational modes
together with their interpretation both in
centrosymmetric $D^{13}_{2h}$ and noncentrosymmetric
$C^7_{2v}$ groups. While the former group explains
naturally the experimental data provided one Raman and
three infrared frequencies not being detected, the
latter group leads to an assumption that 22 of 45
expected Raman and 23 of 37 expected infrared modes were
not detected. Moreover, only three frequencies
(90~cm$^{-1}$, 174 and 951~cm$^{-1}$) coincide within
the experimental precision ($\pm 4$~cm$^{-1}$) in the
sets of Raman and infrared modes corresponding to a
given irreducible representation of the
noncentrosymmetric group, while all the modes should be
both Raman and infrared active in that case. We also
remind once more that we failed to obtain a realistic
set of force constants when carrying out the lattice
dynamics calculations in the assumption of $C^7_{2v}$
noncentrosymmetric space group.

We consider our Raman and infrared data as well as the
results of lattice dynamics calculations to support
strongly the conclusion of the previous structural
study~\cite{smolinski,meetsma} that the space group of
NaV$_2$O$_5 $ above $T_c=35$~K is the
centrosymmetric $D^{13}_{2h}$ rather than
noncentrosymmetric $C^7_{2v}$ group. From the point of
view of $D^{13}_{2h}$ group it is also easy to explain
the results of a recent $^{51}$V-NMR
study~\cite{ohama} that revealed only one vanadium
position at elevated temperatures.

\subsection{Atomic displacements}

As we have already mentioned in the Introduction, the
structure of NaV$_2$O$_5$ looks like the structure of
V$_2$O$_5$ intercalated with Na. The V--O bond lengths
within the vanadium-oxygen layers are close in these two
compounds (see Table~4). The longest bond within the
layer interconnects two V$_2$O$_5$ units in the crystal
unit cell (see Fig.~1b where this bond is indicated as a
dashed line). So, it is of a certain meaning to classify
the $k=0$ crystal vibrations on the basis of internal
vibrations of the V$_2$O$_5$ ``molecule'' ($C_{2v}$
point symmetry group) split into Davydov doublets of the
$D_{2h}$ factor group by an interaction between two
``molecules'' in the crystal unit cell. ($A_g+B_{1u}$),
($B_{2g}+B_{3u}$), ($B_{3g}+B_{2u}$) and ($B_{1g}+A_u$)
Davydov doublets come from $A_1$, $B_1$,
$B_2$ and, correspondingly, $A_2$ vibrations of the
V$_2$O$_5$ ``molecule''. Splittings of these doublets
can be as big as 100~cm$^{-1}$ due to Coulomb
interactions and, in particular, due to interactions
between the adjacent V ions through the common neighbors
(O2 ions) along the chains (see Fig.~1). Many of the
vibrational frequencies of Na V$_2$O$_5$ are close to
those of V$_2$O$_5$~\cite{abello}.

The comparison of our observed vibrational frequencies
with those of V$_2$O$_5$~\cite{abello} and with the
results of our calculations leads to the following
assignment of the vibrational modes of NaV$_2$O$_5$. The
V--O3 stretching modes manifest themselves by two
Davydov doublets: 951~cm$^{-1}$ ($B_{2g}$)~$+$
939~cm$^{-1}$~($B_{3u}$) and 970\,cm$^{-1}$\,($A_g$)~$+$
955~cm$^{-1}$~($B_{1u}$). The Davydov splittings are
relatively small in this case indicating that these
vibrations associated with the strongest bond V--O3 are
really well localized. The mode frequencies are somewhat
lower than the corresponding frequencies in V$_2$O$_5$
(976, 982, 994 and 975~cm$^{-1}$), which is in
accordance with longer V--O3 bonds in NaV$_2$O$_5$ in
comparison with V$_2$O$_5$. The following vibrations are
associated with the O1--V--O3 bending modes:
177~cm$^{-1}$~($A_g$), 162~cm$^{-1}$~($B_{1u}$),
392~cm$^{-1}$~($B_{2g}$), 366~cm$^{-1}$~($B_{3g}$).

The bridging oxygens O1 participate in the bending
V--O1--V vibrations 418~cm$^{-1}$~($B_{3g}$),
367~cm$^{-1}$~($B_{2u}$), 447~cm$^{-1}$~($A_g$),
468~cm$^{-1}$~($B_{1u}$). The stretching V--O1--V
vibration is placed at 420~cm$^{-1}$~($A_g$) and
involves mainly the motion of the V atoms along the $a$-axis.

The modes at 683~cm$^{-1}$~($B_{3g}$ and $B_{1g}$) and
582~cm$^{-1}$ ($B_{2u} $) correspond to the stretching
V--O2 vibrations along the $b$-axis while those at
550~cm$^{-1}$~($B_{2g}$), 533~cm$^{-1}$ ($A_g$) and
526~cm$^{-1}$~($B_{3u}$) --- to the bending ones.

Most of the remaining modes can be described in terms of
external modes of the V$_2$O$_5$ units. Thus, the modes
at 186~cm$^{-1}$ ($B_{2g})$, 169~cm$^{-1}$ ($B_{3g}$)
and 90~cm$^{-1}$~($A_g$), correspond to the relative
translations of the two V$_2$O$_5$ units within the
crystal unit cell along the $a$-, $b$- and,
correspondingly, $c$-axes. As the V$_2$O$_5$ units are
bound together along the $b$-axis, these modes can be
considered as relative translations of the neighboring
(VO$_5$)$_n$ chains. The $B_{1g}$ mode at 174~cm$^{-1}$
(O3 ions move along the $b$-axis) and the $A_g$ mode at
301~cm$^{-1}$ (O2 ions move along the $c$-axis)
correlate with in-plane and, correspondingly,
out-of-plane chain bending vibrations. The $B_{2g}$ mode
at 141~cm$^{-1}$ and $B_{3u}$ mode at 91~cm$^{-1}$ are
connected with the rotation of the chains around the
$b$-axis.

The following modes involve mainly the displacements of
Na atoms: 225~cm$^{-1}$~($B_{2u}$),
251~cm$^{-1}$~($B_{3u}$), 181~cm$^{-1}$~($B_{1u}$).

\subsection{Spectra of electron excitations}

With obtained values of the effective charges, we
estimated the crystal field energies of the 3$d$
electron localized at the V$^{4+}$ ion site. The crystal
field parameters $B_2^0=$1360+2090$G,$
$B_2^1=$2020-1590$G,$ $B_2^2=$820+640$G,$
$B_4^0=$610+1430$G,$ $B_4^1=$-1810-3690$G,$
$B_4^2=$33+144$G,$ $B_4^3=$3650+8180$G,$
$B_4^4=$4070+7420$G$ cm$^{-1}$ for the V$_1,$ V$_3$ sites
were calculated in the framework of the exchange charge
model \cite{malkin} (for the V$_2$, V$_4$ sites $B_p^1$
and $B_p^3$ parameters change signs, the first terms
correspond to point charge contributions, the Stevens
normalization is used). The scaling factor $G$
determines the strength of the exchange charge field. We
estimated this phenomenological parameter of the model
($G=4$) by fitting the total crystal field splitting to
the width of the V--3$d$ bands presented in Ref.~\cite{smolinski}. In
this case, the effective crystal field provides the following energy
level pattern of the V$^{4+}$ ion : 0 ($A''$), 1.10 ($A'$), 1.18
($A''$), 3.39 ($A'$) and 4.78 ($A'$)~eV (irreducible representations
of the $C_s$ point group, corresponding to the space symmetry of the
electron wave function, are given in brackets, additional shifts of
the crystal field levels due to the spin-orbit interaction and the
electrostatic field of a hole at the neighboring
vanadium site are less than 0.025 eV). The ground state
wave function is the $d_{xy}$ orbital with small
admixture of the $d_{yz}$ orbital as it has been already
pointed earlier \cite{smolinski}, and the sequence of
the excited states is in agreement with results of the
band structure calculations as well \cite{smolinski}.

The strong absorption of light (${\bf k}||{\bf c}$) with
the ${\bf E}||{\bf a}$ as well as with ${\bf E}||{\bf
b}$ polarizations was observed in the region of 
1.0--1.2~eV~\cite{we}. Both magnetic dipole and induced electric
dipole $d-d$ transitions in the odd crystal field are
allowed between the $A''$ states for ${\bf E}||{\bf a}$, and between
the $A''$ and $A'$ states for ${\bf E}||{\bf b}$. Thus, in accordance
with the results on the crystal field energies given above, the
observed broad optical bands in the region of 1~eV can be interpreted
as phonon assisted $d-d$ transitions without any additional
suppositions about the broken symmetry between legs of vanadium
ladders \cite{Damasc}.

The next step towards the detailed description of the spectra of
electron excitations is to construct molecular orbitals for the
$[$V$_2$O$]^{7+}$ ``molecule'' having $C_{2v}$ point symmetry, from the
vanadium $d$-orbitals and oxygen $p$-orbitals. The vanadium ground
state wave function $d_{xy}$ yields the non-bonding $a_2$ orbital as
well as bonding and antibonding $b_2$ molecular orbitals, namely, \\
$a_2:   [d_{xy}(L)+d_{xy}(R)]$ \\
$b_2:   [d_{xy}(L)-d_{xy}(R)] + p_y$ \\
$b_2^*: [d_{xy}(L)-d_{xy}(R)] - p_y $. \\
Here, $a_2$, $b_2$ denote irreducible representations of the $C_{2v}$
point group, the asterisk refers to an antibonding orbital, $L$ and $R$
stand for the vanadium sites at the left and, correspondingly, right sides
of a ladder rung. The highest filled orbital being $a_2$, the
$a_2\rightarrow b_2^*$ electronic transition allowed in ${\bf E}||{\bf a}$
polarization could account for the low-frequency absorption band observed
just in this polarization. Quantum-chemical calculations are necessary to
verify this qualitative interpretation.

\subsection{Fano resonances with a continuum}

The asymmetric lineshapes of the infrared active modes
at 91, 150, 939~cm$^{-1}$ in ${\bf E}||{\bf a}$
polarization bring into evidence a strong interference
between these modes and a continuum observed just in
this polarization. This interpretation is supported by
the fact that the spectral line near 91~cm$^{-1}$
becomes perfectly symmetric when the continuum
absorption vanishes in this spectral range below the
phase transition temperature $T_c=35$~K. In our earlier
work \cite{LJETP} we argued that these changes are
connected with the opening of a gap in the magnetic
excitation spectrum at $T_c$, the observed continuum
being due to two-magnon absorption.

However, such a straightforward interpretation is no more valid in
the case of the space group $D_{2h}^{13}$.  It should be revised,
taking into account possible electronic excitations in this frequency
range, as discussed in the previous section, and a charge ordering at
the transition temperature.

\subsection{Higher order infrared vibrational spectra}

Two- and three-phonon absorption occurs due to
anharmonicity of crystal vibrations. It is continuous,
displaying singularities corresponding to critical
points of the Brillouin zone. Leaving the detailed
analysis of multiphonon bands to a future publication,
we discuss here only sharp lines observed in ${\bf
E}||{\bf c}$ absorbance spectrum (Fig.~5). They are listed in Table 5
together with their tentative assignment, using symmetry allowed
combinations of $\Gamma$-point phonons observed in our first-order
spectra. The coincidence of the observed and combinational
frequencies is within the precision of our measurements.

The strongest narrow peak at 1930~cm$^{-1}$ corresponds, according to
this assignment, to a sum of the components of the Davydov doublet
originating from the V--O3 stretching vibration. This stretching mode
is well localized which results in its small dispersion over the
Brillouin zone, thus delivering a narrow two-phonon band, in
accordance with the experimental observation.

\section{Summary}

We have performed a thorough spectroscopic study of far infrared
reflection and transmission as well as Raman scattering of
$\alpha'$-NaV$_2$O$_5$ single crystals in the high temperature phase
(above $T_c=35$~K).  Far infrared spectra were obtained for ${\bf
E}||{\bf a}$, ${\bf E}||{\bf b}$ and ${\bf E}||{\bf c}$ polarizations
of incident light.  Diagonal $(aa)$, $(bb)$, $(cc)$ as well as
nondiagonal $(ab)$, $(bc)$, $(ac)$ components of the Raman scattering
tensor were investigated. We report five infrared active modes in
${\bf E}||{\bf a}$ polarization, four~--- in ${\bf E}||{\bf b}$
polarization and six~--- in ${\bf E}||{\bf c}$ polarization. Eight
Raman active modes have been detected for parallel polarizations of
incident and scattered light $(aa)$, $(bb)$, $(cc)$. The $(ab)$,
$(ac)$ and $(bc)$ Raman geometries delivered three, seven and,
correspondingly, five modes. These results are in a much better
agreement with the recently proposed  centrosymmetric space group
$D_{2h}^{13}$ ($Pmmn$) for the high-temperature phase of NaV$_2$O$_5$
than with earlier accepted noncentrosymmetric space group $C_{2v}^7$
($Pmn2_1$). We have also performed the lattice dynamics calculations,
in the framework of the rigid ion model, for both proposed in the
literature structures of NaV$_2$O$_5$.  We failed to obtain any
physically well-grounded set of parameters providing a stable
$C_{2v}^7$ lattice structure. Thus, our infrared and Raman
experimental data, as well as the results of lattice dynamics
calculations support strongly the conclusion of the previous
structural study~\cite{smolinski,meetsma} that the space group of
NaV$_2$O$_5$ above $T_c=35$~K is the centrosymmetric $D^{13}_{2h}$
rather than noncentrosymmetric $C^7_{2v}$ group.

This conclusion leads to important physical consequences. In
particular, it requires to revise the interpretation of
one-dimensional magnetic properties of NaV$_2$O$_5$ and of the phase
transition at 35~K considered earlier as an ordinary spin-Peierls
transition.  An interpretation of the earlier observed broad bands
in the near and far infrared absorption~\cite{we,LJETP} needs to be
reconsidered too.

Using the values of effective charges obtained in the process of
lattice dynamics calculations and fitting the total crystal field
splitting to the width of the V--$3d$ bands~\cite{backman}, we
estimated the crystal field energies of the $3d$ electron localized
at the vanadium site. It follows from this estimate that the observed
near infrared broad band absorption of NaV$_2$O$_5$~\cite{we} can be
interpreted as phonon assisted $d-d$ transitions. We speculate that
the far infrared ${\bf E}||{\bf a}$ polarized absorption continuum
might be associated with electron excitations of $[$V$_2$O$]^{7+}$
rungs in the crystal field of $C_{2v}$ symmetry.

Strongly asymmetric spectral lines observed in ${\bf E}||{\bf a}$
absorbance spectra of NaV$_2$O$_5$ bring into evidence a strong
interference between relatively narrow phonon lines and the
underlying continuum. This points to an interaction between crystal
vibrations and magnetic or electronic excitations. The detailed
physical interpretation of the observed phenomenon depends on the
nature of the far infrared ${\bf E}||{\bf a}$ polarized continuum
which requires a special investigation.

In conclusion, we reported also some preliminary results on higher
order vibrational spectra of NaV$_2$O$_5$ appearing due to
anharmonicity of lattice vibrations.

\acknowledgments
We are grateful to A. I. Smirnov for checking the samples by ESR
measurements. We acknowledge A. N. Vasil'ev for stimulating
discussions and G. N. Zhizhin for a constant support of this
research. This work was made possible in part by the Grant
No.~98-02-17620 from the Russian Fund for Basic Research.

\newpage

\section*{Figure captions}

\begin{figure}
\caption{The structure of NaV$_2$O$_5$. (a) A
stereometric projection. Oxygen and vanadium atoms are
at the corners of and, correspondingly, inside the
pyramids. Sodium atoms are presented as balls. (b) $ab$
projection. Apical oxygen O3 atoms (situated above or
below the corresponding V atoms) are not shown. A dashed
line indicates the longest V--O2 bond
(0.199~nm).
$ab$-projection of the crystal unit cell is shown by a
thin solid line.}
\end{figure}

\begin{figure}
\caption{Room temperature far-infrared reflectivity
spectra of NaV$_2$O$_5$. Open circles represent
experimental data. Solid lines are the result of the
fitting (see the text).}
\end{figure}

\begin{figure}
\caption{Absorption coefficient $\alpha$ in the region of low
frequency absorption bands at room temperature. An arrow indicates a
Fano-type resonance. It is shown separately in the inset. Open
circles represent experimental data. Solid line in the main figure
was calculated using the parameters obtained from the fitting of the
reflectance spectrum. Solid line in the inset is a result of the
fitting by the expression~(5)
with $\alpha_B(\omega)$ shown as a dashed line.}
\end{figure}

\begin{figure}
\caption{Fano resonance near 91~cm$^{-1}$ at the
temperature of 40~K (open circles) and its fitting by
the equation (5) with parameters $\omega_r=90.7$~cm$^{-1}$,
$\gamma=0.2$~cm$^{-1}$, $q=-1.0$,
$\alpha_B(\omega_2)=270$~cm$^{-1}$,
$\alpha_0/\alpha_B=0.3$ (solid line). The temperature
dependence of the Fano parameter $q$ is given in the
left inset. The right inset presents the absorption
spectrum in the vicinity of 91~cm$^{-1}$ line at 6~K
taken with the resolution of 0.05~cm$^{-1}$ (open
circles) and its fitting by the Lorenzian with a
FWHM=0.10~cm$^{-1}$.}
\end{figure}

\begin{figure}
\caption{Absorbance spectrum of NaV$_2$O$_5$ in the
region of multiphonon bands at room temperature.}
\end{figure}

\begin{figure}
\caption{Room temperature Raman spectra of NaV$_2$O$_5$.
Asterisks mark $A_g$ lines seen in $B_{ig}$
($i=1, 2, 3$) spectra.}
\end{figure}

\begin{figure}
\caption{Room temperature Raman spectra of
Na$_{1-x}$V$_2$O$_5$ for different $x$.}
\end{figure}

\begin{figure}
\caption{Decomposition of the room temperature
$c(aa)\bar{c}$ Raman spectrum of NaV$_2$O$_5$ into
separate contours (dashed lines). The sum of these
contours shown by a solid line approximates the
experimental spectrum (circles) rather well.}
\end{figure}
\newpage
\begin{table}[t]
\caption{Infrared active vibrational modes and dielectric
constants of NaV$_2$O$_5$ (all frequencies are in cm$^{-1}$) }
\begin{tabular}{c|rrrrrr|rrrrrrr|rr}

~ & \multicolumn{13}{c|}{Observed} & \multicolumn{2}{c}{Calculated} \\
Polarization, & \multicolumn{13}{c|}{~ } & \multicolumn{2}{c}{($Pmmn$)} \\
\cline{2-16}
mode & \multicolumn{6}{c|}{Transmission} &
  \multicolumn{7}{c|}{Reflection} & ~ & ~ \\
\cline{2-14}
symmetry & \multicolumn{2}{c}{$T=40$~K} & \multicolumn{4}{c|}{$T=300$~K} &
      \multicolumn{7}{c|}{$T=300$~K} & ~ & ~ \\
~ & $\omega_{TO}$ & $\gamma_{TO}$ & $\omega_{TO}$ & $\gamma_{TO}$ &
           $\varepsilon_{\omega_1}$\tablenotemark[1]   & $\varepsilon_{\omega_2}$\tablenotemark[1] &
           $\omega_{TO}$ & $\gamma_{TO}$ & $\omega_{LO}$ & $\gamma_{LO}$ &
           $f\cdot 10^3$ & $\varepsilon_{\infty}$ & $\varepsilon_0$ &
           $\omega_{TO}$ & $\omega_{LO}$ \\
\hline
${\bf E}||{\bf c}$ &  ~  & ~ & ~ & ~ & ~ & 7.5 & 162 & 5.2 & 165 & 5.7 & 45 & 3.9 & 7.7 & 216 & 219 \\
$B_{1u}(Pmmn)$     &  ~  & ~ & ~ & ~ & ~ & $\pm0.2$ & 179 & 8.4 & 212 & 8.3 & 130 & ~ & ~ & 232 & 256 \\
or                 &  ~  & ~ & ~ & ~ & ~ & ~ & --- & ~   &   ~ &   ~ & ~   & ~ & ~ & 298 & 298 \\
$B_2(Pmn2_1)$      &  ~  & ~ & ~ & ~ & ~ & ~ & 468 & 38.0& 483 & 38.0& 23  & ~ & ~ & 430 & 430 \\
                   &  ~  & ~ & ~ & ~ & ~ & ~ & 591 &119.9& 594 &119.0& 6.8 & ~ & ~ & 589 & 690 \\
                   &  ~  & ~ & ~ & ~ & ~ & ~ & 760 & 59.4& 762 & 59.3& 1.6 & ~ & ~ & 691 & 716 \\
                   &  ~  & ~ & ~ & ~ & ~ & ~ & 955 & 2.5 &1017 & 3.0 & 39  & ~ & ~ & 961 &1036 \\[5mm]
${\bf E}||{\bf b}$ & 178 &  4  & 175 & 12 & 5.2 & 10.2  & 175 & 8.3 &180 & 8.4 & 39 & 4.9 & 9.5 & 141 & 173 \\
$B_{2u}(Pmmn)$     & 225 &  1  & --- & ~  & $\pm0.2$ & $\pm0.2$  &  ~  &  ~  &  ~  & ~   &  ~ & ~ & ~ & 240 & 266 \\
or                 & 371 &  ~  & 367 & 16 & ~ & ~  & 365 & 12.8& 378 & 13.3& 60 & ~ & ~ & 388 & 483 \\
$B_1(Pmn2_1)$      & 594 & 13  & 582 & ~ & ~ &  ~  & 584 & 29.5& 769 & 29.0& 271& ~ & ~ & 578 & 747 \\[5mm]
${\bf E}||{\bf a}$ & 91\tablenotemark[3] & ~&  ---  & ~ & 9.6 &  15.0  & ~   & ~   & ~   &  ~  &  ~ & 7.7& 15.8& 111 & 126 \\
$B_{3u}(Pmmn)$     & 140\tablenotemark[2] & ~ & 145\tablenotemark[2] & ~ & $\pm0.3$ & $\pm0.6$ & 153 & 33.0& 154 & 34.1& 46 & ~ & ~ & 130 & 177 \\
or                 & 254 &  ~  & 251 &  ~  &~ & ~ & 251 & 7.3 & 252 & 8.2 &  9 & ~ & ~ & 227 & 276 \\
$A_1(Pmn2_1)$                   & --- &  ~  & ~   &  ~  & ~ & ~ & --- & ~   & ~   & ~   & ~  & ~ & ~ & 493 & 534 \\
                   & 531 & 18  & 526 & 53  & ~ & ~ & 525 & 39.5& 621 & 52.4& 208& ~ & ~ & 538 & 653 \\
                   & --- &  ~  & ~   &  ~  & ~ & ~ & --- & ~   & ~   & ~   & ~  & ~ & ~ & 742 & 808 \\
                   & ~   &  ~  & 939\tablenotemark[4] & ~ & ~ &  ~  & ~   & ~   & ~   & ~   & ~  & ~ & ~ & 955 & 957 \\
\end{tabular}
\tablenotetext[1]{$\omega_1=3200$~cm$^{-1}$, $\omega_2=40$~cm$^{-1}$.}
\tablenotetext[2]{Asymmetric line.}
\tablenotetext[3]{Fano-type resonance: $\omega_r=90.7$~cm$^{-1}$,
                 $\gamma=0.2$~cm$^{-1}$, $q=-1.0$, $\alpha_0/\alpha_B=0.3$.}
\tablenotetext[4]{Fano-type resonance: $\omega_r=939$~cm$^{-1}$,
                      $\gamma=1.0$~cm$^{-1}$, $q=1.1$, $\alpha_0/\alpha_B=0.2$.}
\end{table}
\begin{table}[h]
\caption{Room temperature Raman frequencies (cm$^{-1}$) for NaV$_2$O$_5$}
\begin{tabular}{lc|cc|cc|cc}
\multicolumn{2}{c|}{$aa, bb, cc$}&
\multicolumn{2}{c|}{$ab$}&
\multicolumn{2}{c|}{$ac$}&
\multicolumn{2}{c}{$bc$} \\
\multicolumn{2}{c|}{$A_{1g}(Pmmn)$} &
\multicolumn{2}{c|}{$B_{1g}(Pmmn)$} &
\multicolumn{2}{c|}{$B_{2g}(Pmmn)$} &
\multicolumn{2}{c}{$B_{3g}(Pmmn)$}\\
\multicolumn{2}{c|}{or $A_1(Pmn2_1)$} &
\multicolumn{2}{c|}{or $B_1(Pmn2_1)$} &
\multicolumn{2}{c|}{or $B_2(Pmn2_1)$} &
\multicolumn{2}{c}{or $A_2(Pmn2_1)$} \\
\hline
Observed & Calc.($Pmmn$) & Observed & Calc.($Pmmn$) & 
Observed & Calc.($Pmmn$) & Observed & Calc.($Pmmn$) \\
\hline
90 $(cc, aa)$      &  91 & 174  & 191 & 141 & 129 & 169 & 149  \\
177 $(aa, bb, cc)$ & 226 & 295  & 288 & 186 & 193 & 257 & 239  \\
233 $(bb)$         & 319 & 683  & 679 & 225 & 296 & 366 & 262  \\
301 $(aa, bb)$     & 362 &  ~   &  ~  & 392 & 332 & 418 & 396  \\
420 $(bb, cc)$     & 439 &  ~   &  ~  & 429 & 410 & 683 & 685  \\
447 $(aa)$         & 518 &  ~   &  ~  & 550 & 619 &  ~  &  ~   \\
533 $(aa, bb)$     & 626 &  ~   &  ~  & --- & 798 &  ~  &  ~   \\
970 $(cc, aa)$     & 964 &  ~   &  ~  & 951 & 961 &  ~  &  ~   
\end{tabular}
\end{table}
\begin{table}[h]
\caption{Comparison of experimentally observed Raman and infrared modes with
the expected ones within centrosymmetric $D_{2h}^{13}$ and
noncentrosymmetric $C_{2v}^7$ space groups (mode frequencies are in cm$^{-1}$)}
\begin{tabular}{lllllllllcl}

$Pmmn(D_{2h}^{13})$ & \multicolumn{8}{c}{Observed modes} &
\multicolumn{2}{c}{$Pmn2_1(C_{2v}^7)$} \\[2mm]
\hline
\hline
$8A_g (aa,bb,cc)$ \raisebox{5mm}{~}&
                  90  & 177 & 233 & 301 & 420 & 447 & 533 & 970 &
$\backslash$ & \raisebox{-1.7ex}{$15A_1 (aa,bb,cc; {\bf E}||{\bf a})$} \\
$7B_{3u} ({\bf E}||{\bf a})$ &
                  91\tablenotemark[1]
                  & 145 & 251 & 526 & 939 & ~   & ~   & ~ & / & ~ \\[2mm]
\hline
$3B_{1g} (ab)$ \raisebox{5mm}{~}&
174 & 295 & 683 & ~ & ~ & ~ & ~ & ~ &
$\backslash$ & \raisebox{-1.7ex}{$7B_1$ $(ab; {\bf E}||{\bf b})$}\\
$4B_{2u} ({\bf E}||{\bf b})$ &
                   175 & 225\tablenotemark[1]
                       & 367 & 582 & ~ & ~ & ~ & ~ & / &~ \\[2mm]
\hline
$8B_{2g} (ac)$ \raisebox{5mm}{~}&
141 & 186 & 225 & 392 & 429 & 550 & 951 & ~ &
$\backslash$ & \raisebox{-1.7ex}{$15B_2$ $(ac; {\bf E}||{\bf c})$} \\
$7B_{1u} ({\bf E}||{\bf c})$ &
                  162 & 179 & 468 & 591 & 760 & 955 & ~ & ~ & / &~ \\[2mm]
\hline
$5B_{3g} (bc)$ \raisebox{5mm}{~}&
169 & 257 & 366 & 418 & 683 & ~ & ~ & ~ &
$\backslash$ & \raisebox{-1.7ex}{$8A_2$ $(bc)$} \\
$3A_u$ &~ &~ &~ &~ &~ &~ &~ & ~ & / & \\[2mm]
\end{tabular}
\tablenotetext[1]{Observed below 200~K.}
\end{table}
\begin{table}[h]
\caption{The bond lengths (nm) in NaV$_2$O$_5$ and V$_2$O$_5$}
\begin{tabular}{ l l l }

Bond &  NaV$_2$O$_5$ \cite{smolinski} & V$_2$O$_5$ \cite{abello} \\
\hline
V--O3             & 0.161 & 0.158  \\
V--O1             & 0.183 & 0.177  \\
(V--O2)$\times 2$ & 0.192 & 0.188  \\
V--O2$'$          & 0.199 & 0.202  \\
V--O3$'$          & 0.318 & 0.278
\end{tabular}
\end{table}
\begin{table}[h]
\caption{Multiphonon bands in NaV$_2$O$_5$
observed in ${\bf E}||{\bf c}$ polarization}
\begin{tabular}{c|lll}
Observed bands, cm$^{-1}$ & \multicolumn{3}{c}{Combination of phonons,
cm$^{-1}$} \\
\hline
1072 ($B_{1u}$) & 550 ($B_{2g}$) &+ 526 ($B_{3u}$) &= 1076 ($B_{1u}$) \\
1270 ($B_{1u}$) & 683 ($B_{3g}$) &+ 582 ($B_{2u}$) &= 1265 ($B_{1u}$) \\
1930 ($B_{1u}$) & 970 ($A_g$)    &+ 955 ($B_{1u}$) &= 1925 ($B_{1u}$) \\
2858 ($B_{1u}$) & $3\times955$ ($B_{1u}$) & ~ &= 2865 ($B_{1u}$) \\
2901 ($B_{1u}$) & $2\times970$ ($A_g$) &+ 955 ($B_{1u}$) &= 2895 ($B_{1u}$) \\
\end{tabular}
\end{table}
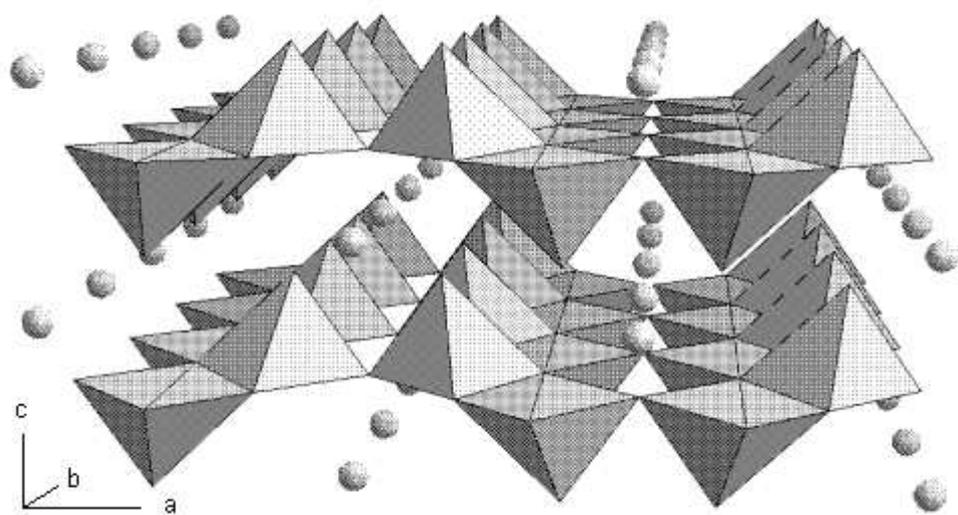
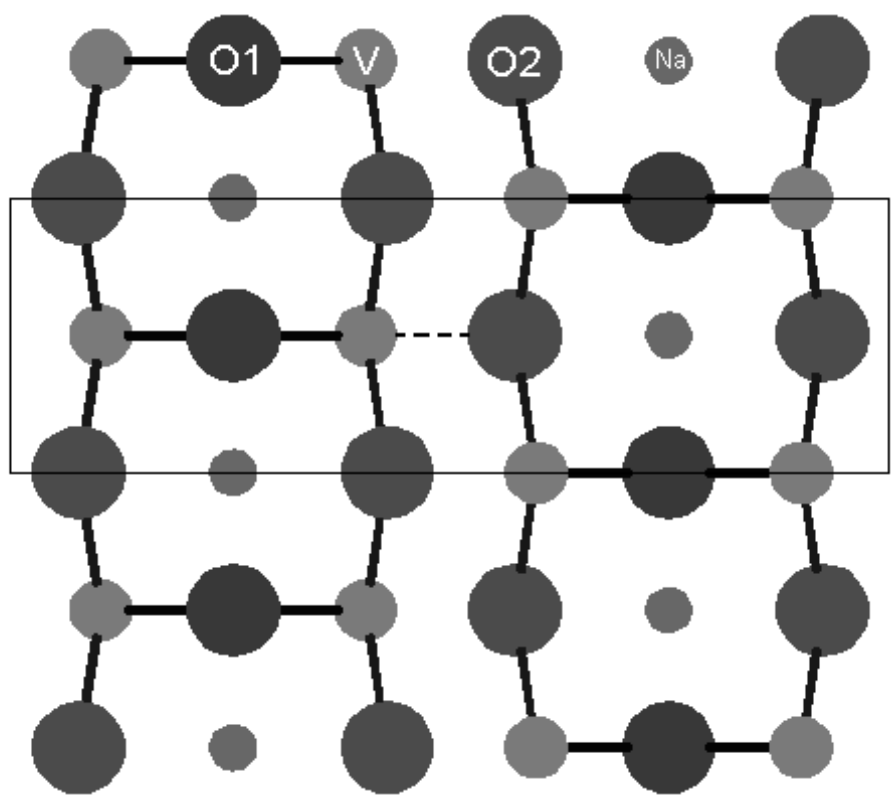
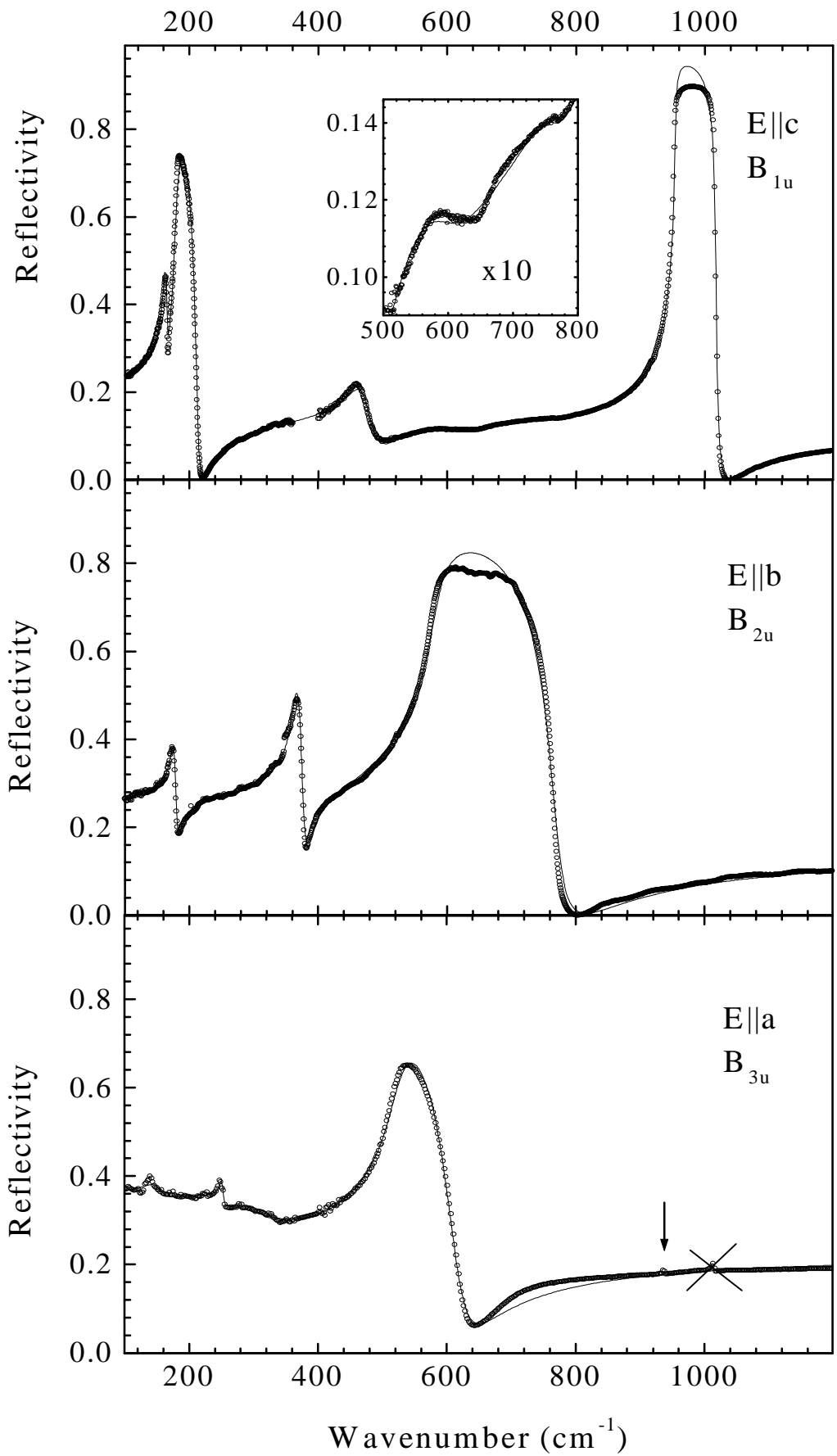
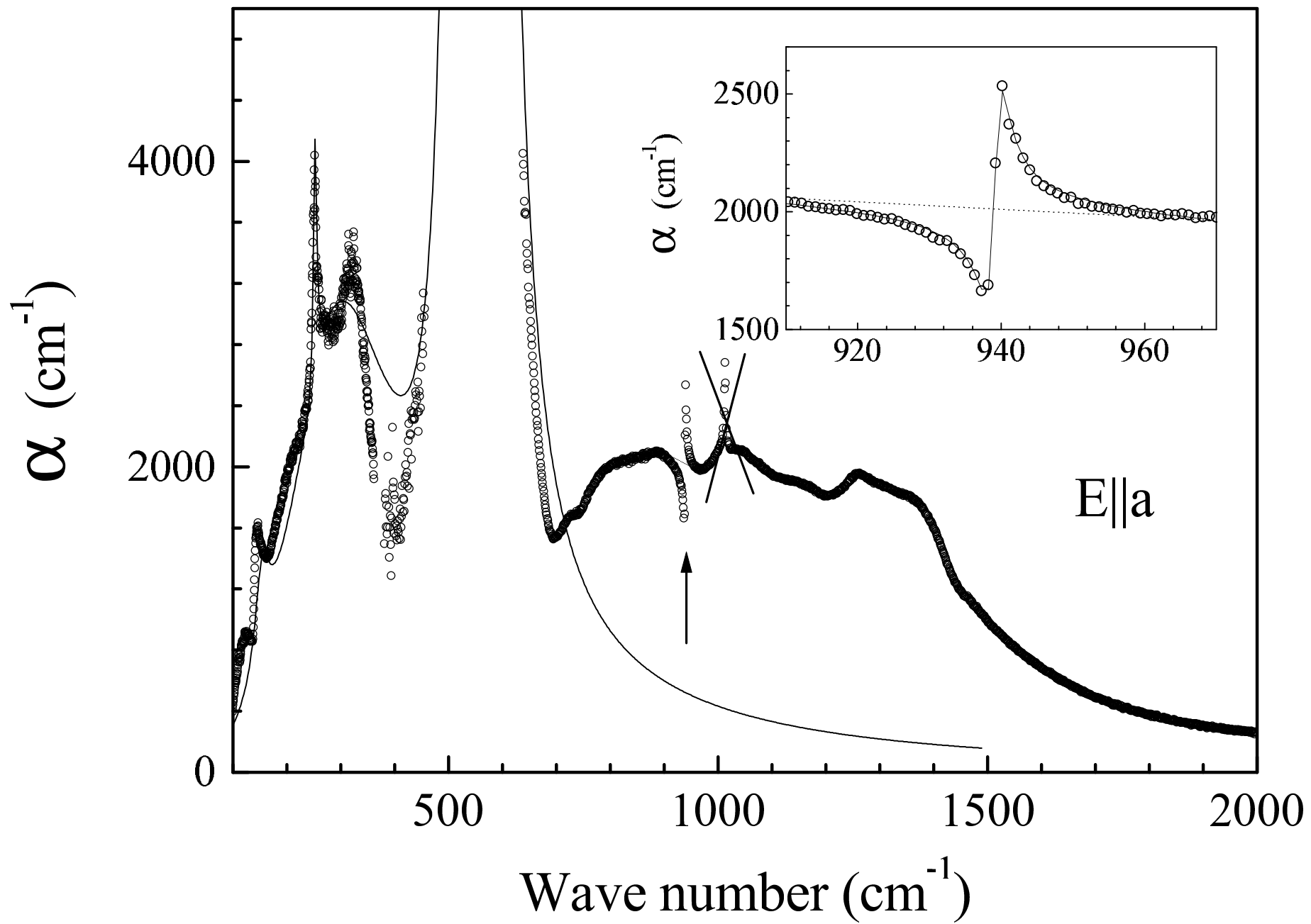
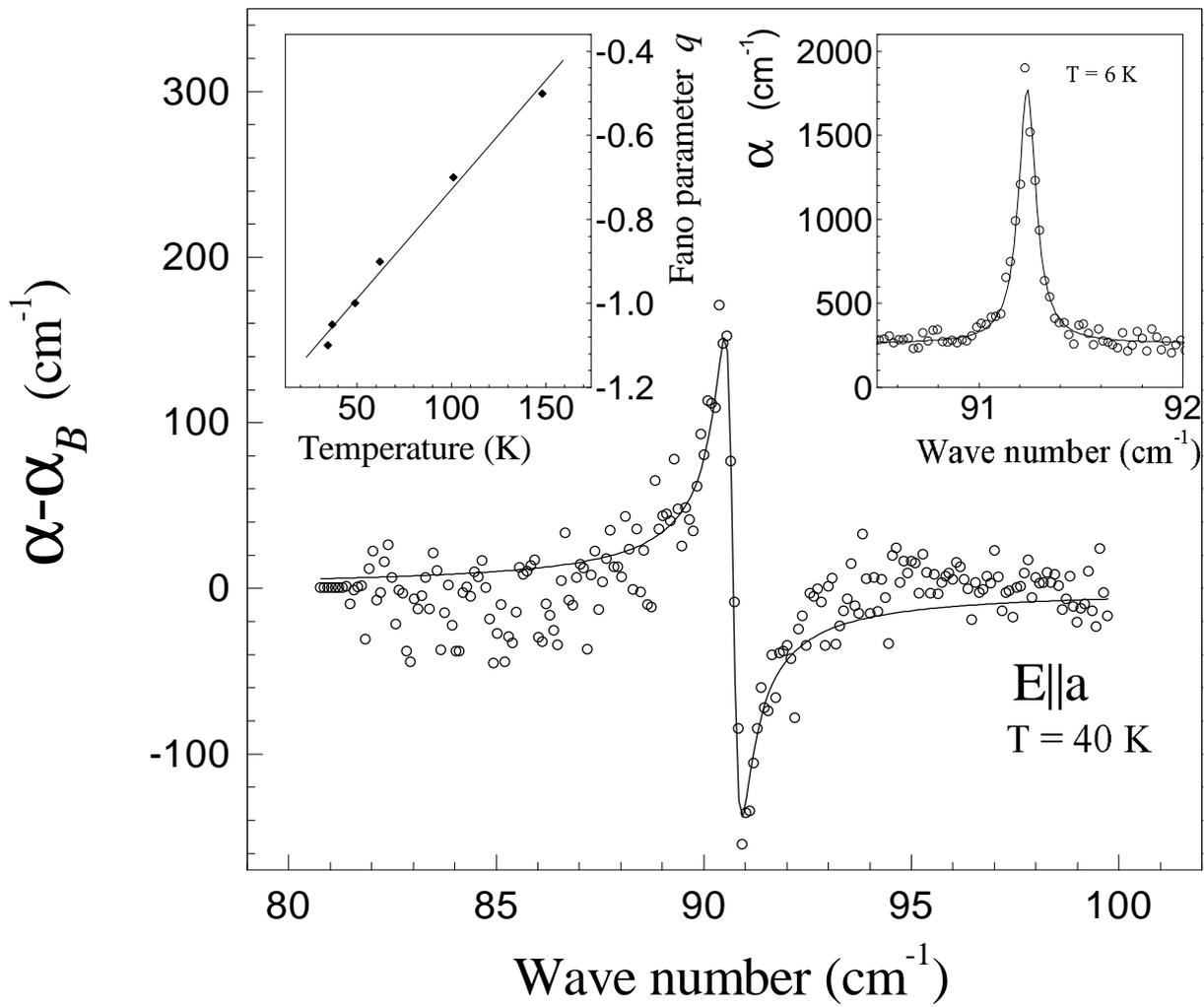
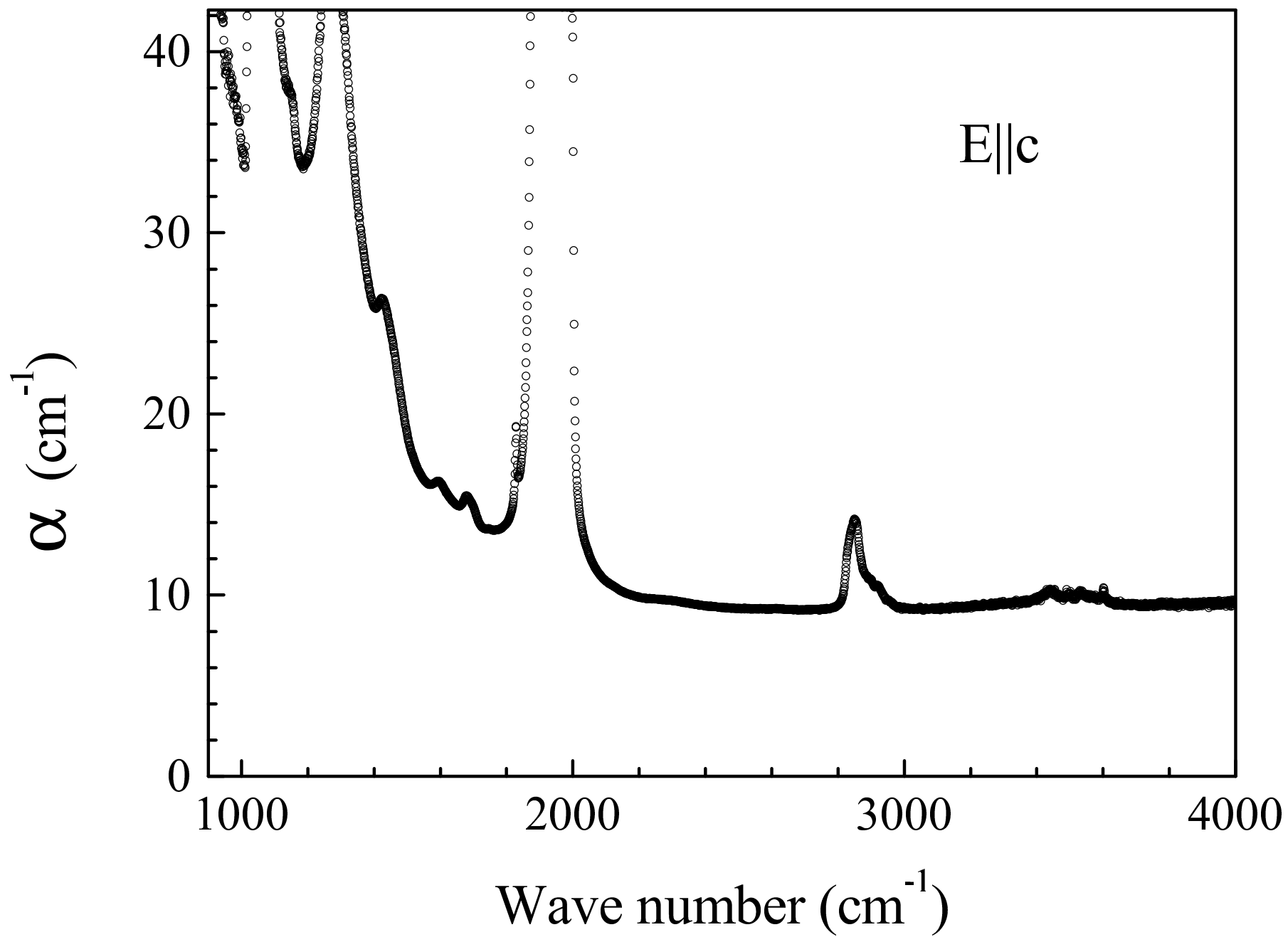
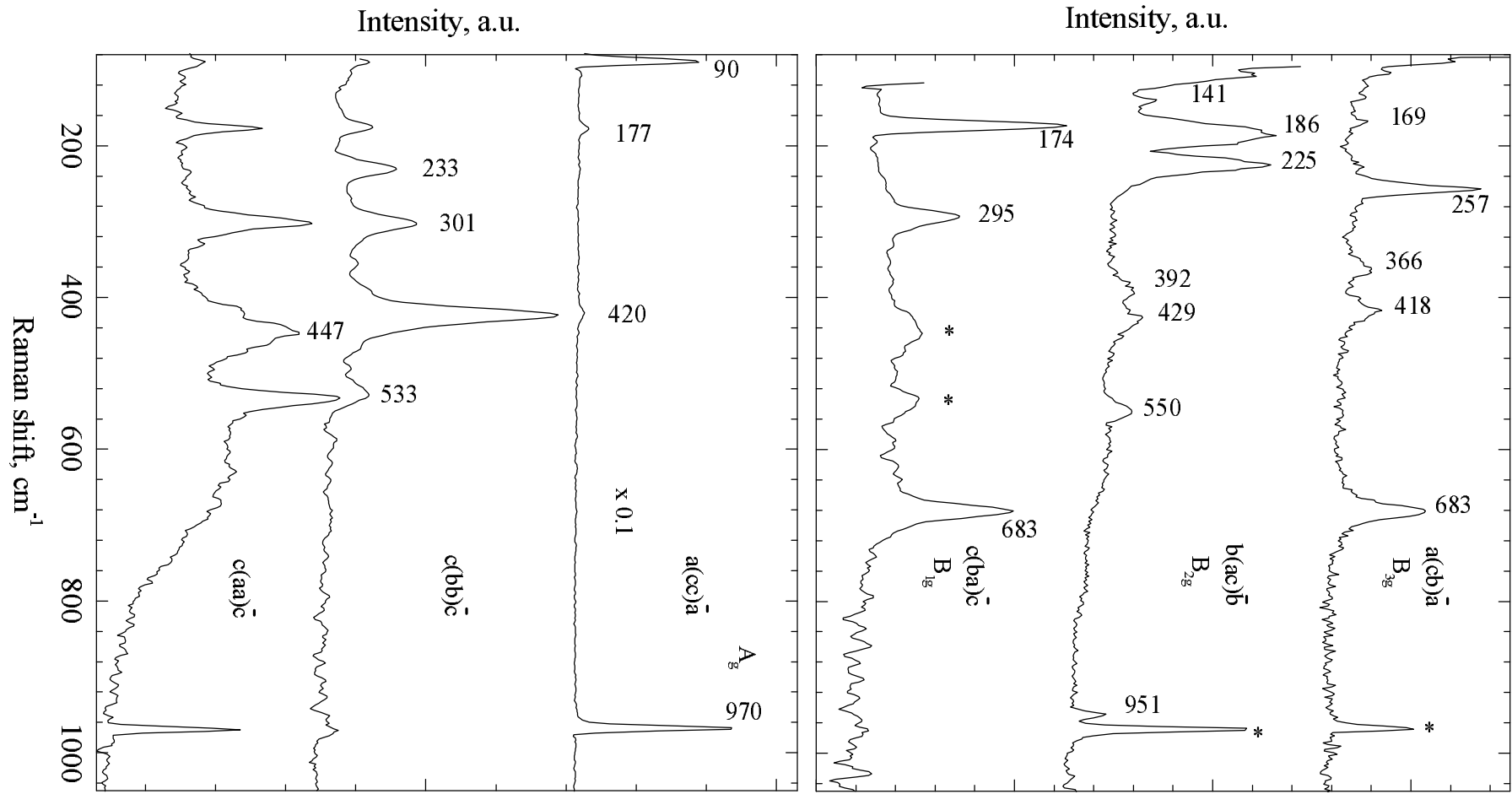
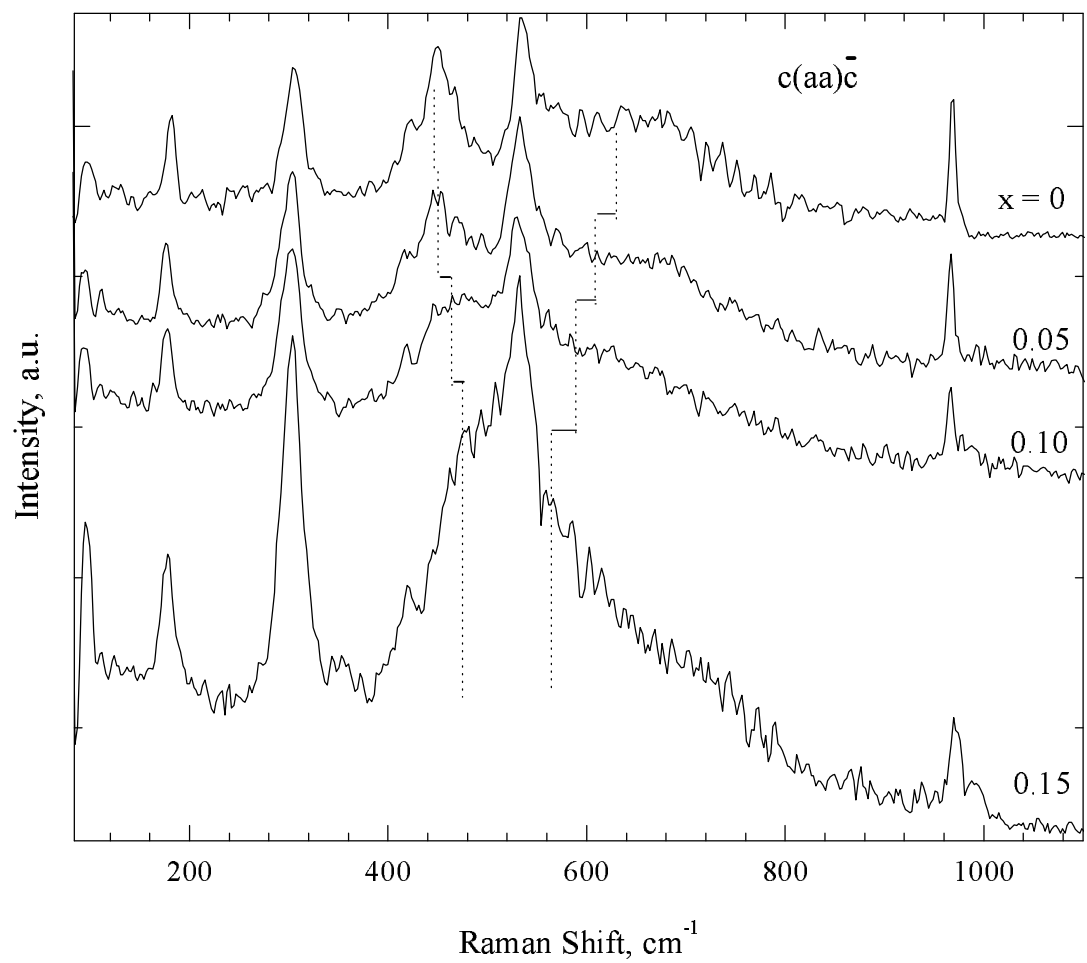
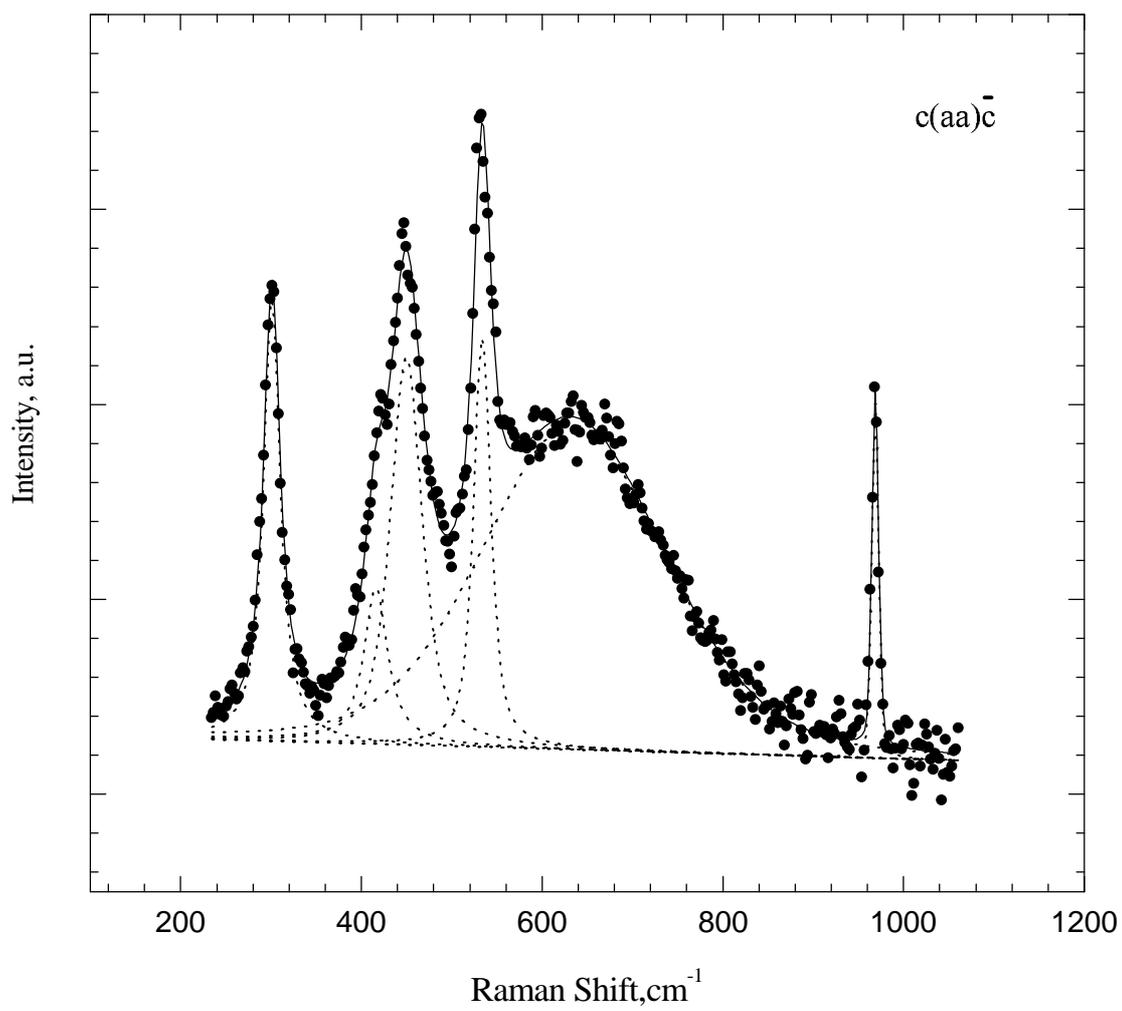
\end{document}